\documentclass{article}
\usepackage{spconf,amsmath,graphicx}
\usepackage{multirow}
\usepackage{booktabs}
\usepackage{amssymb}
\usepackage{bm}

\usepackage{enumitem}
\setlist{nosep, leftmargin=14pt}

\usepackage{mwe} 

\title{Spectral brain graph neural network for prediction of anxiety in children with Autism Spectrum Disorder\\ }

\address{
$^{1}$ Department of Biomedical Engineering, Yale University, USA\\
$^{2}$ Child Study Center, Yale University School of Medicine, USA \\}
\begin{document}
%
\maketitle

\begin{abstract}
Children with Autism Spectrum Disorder~(ASD) frequently exhibit comorbid anxiety, which contributes to impairment and requires treatment. Therefore, it is critical to investigate co-occurring autism and anxiety with functional imaging tools to understand the brain mechanisms of this comorbidity. Multidimensional Anxiety Scale for Children, 2nd edition~(MASC-2) score is a common tool to evaluate the daily anxiety level in autistic children. Predicting MASC-2 score with Functional Magnetic Resonance Imaging~(fMRI) data will help gain more insights into the brain functional networks of children with ASD complicated by anxiety. However, most of the current graph neural network~(GNN) studies using fMRI only focus on graph operations but ignore the spectral features. In this paper, we explored the feasibility of using spectral features to predict the MASC-2 total scores. We proposed SpectBGNN, a graph-based network, which uses spectral features and integrates graph spectral filtering layers to extract hidden information. We experimented with multiple spectral analysis algorithms and compared the performance of the SpectBGNN model with CPM, GAT, and BrainGNN on a dataset consisting of 26 typically developing and 70 ASD children with 5-fold cross-validation. We showed that among all spectral analysis algorithms tested, using the Fast Fourier Transform~(FFT) or Welch's Power Spectrum Density (PSD) as node features performs significantly better than correlation features, and adding the graph spectral filtering layer significantly increases the network's performance. 

\end{abstract}
\section{Introduction}
\label{sec:intro}

\begin{figure*}[!t]
\centering
 \includegraphics[width=\linewidth]{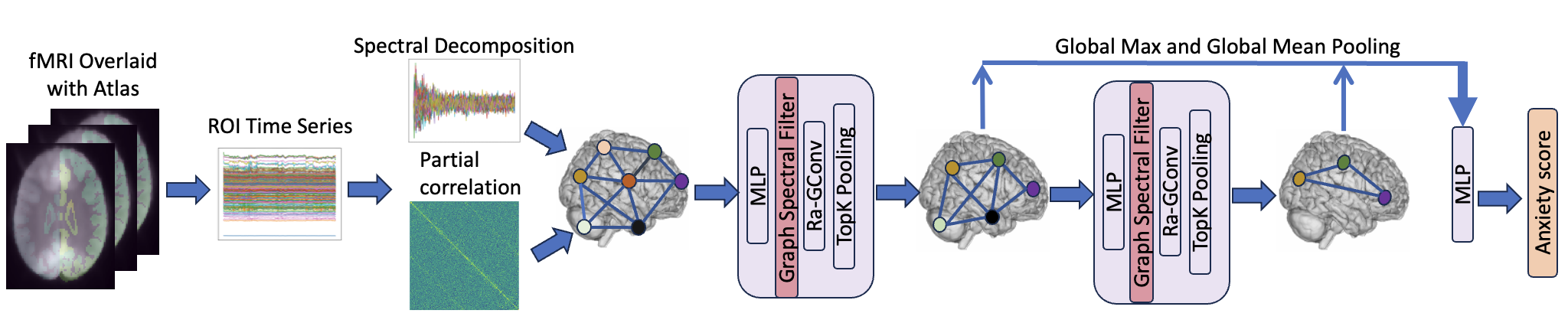}
  \caption{Architecture of the SpectBGNN. Partial correlation matrices are used as edge features while spectral information is used as node features. The graph node features are first fed into an MLP, and then passed into the Graph Spectral Filter to extract useful frequency bands. This is followed by Ra-GConv \cite{braingnn-short}, which updates the graph node features with aggregation of different kernels based on each node's and their neighbors' soft community assignment. The topK pooling layer reduces the number of nodes in the graph by keeping the top half of the important nodes. The global max and global mean pooling of the node features are extracted from this graph. This process is repeated once and the global max and mean pooling results are concatenated with those of the previous graph to feed in the final readout 2-layer MLP to predict the MASC-2 score.}
  \label{fig}
\end{figure*}

Autism Spectrum Disorder (ASD) is a neurodevelopmental condition characterized by challenges in social interaction, communication, and repetitive behaviors \cite{ASD-short}. Studies have also shown that 40\% of ASD patients are often diagnosed with comorbid anxiety \cite{anxiety_autism_comorbidity-short}. Comorbid anxiety presents more challenges to therapy as it causes distress and requires additional treatment. To quantify the anxiety level of children, the Multidimensional Anxiety Scale for Children, 2nd edition~(MASC-2) score \cite{MASC2-short} is specifically designed for adolescents and children, which comprises four major factors: (1) physical symptoms, (2) social anxiety, (3) harm avoidance, and (4) separation anxiety. Moreover, Functional Magnetic Resonance Imaging (fMRI) has become a powerful tool in neuroimaging, providing insights into the brain's functional connectivity. Therefore, predicting MASC-2 total scores from fMRI data offers insights into the brain's network related to anxiety and helps assess the outcome of psychiatric treatment for children.  \\

Integrating graph neural networks (GNNs) with fMRI data holds promise for uncovering nuanced functional connectivity alterations in the brains of ASD-affected children. While various studies have demonstrated the effectiveness of GNNs in analyzing neuroimaging data, showcasing their ability to identify subtle patterns in the functional connectivity (FC) that may elude traditional analysis methods \cite{braingnn-short,fMRI_regGNN-short,gnn_task-fmri-short}, very few studies have explored the impact of substituting the Pearson's correlation with spectral features on the model's performance. 
\\

Fourier analysis methods have been applied widely in signal processing fields and neural network computations. Examples of spectral information in biomedical-related tasks include but are not limited to electrocardiogram signal classification \cite{fft_ecg-short} and electroencephalogram signal classification \cite{fft_eeg-short}. Some studies have already shown that spectral features could better capture the connectivity information that otherwise would not be captured by Pearson's correlation \cite{spectral_feautres}\cite{rethinking_connectivity}. In the pursuit of a comprehensive analysis, our study endeavors to evaluate the graph network's performance when supplying the node features with either frequency domain information or Pearson correlations \cite{braingnn-short}. This exploration necessitates a diverse set of spectral analysis methodologies, encompassing Fast Fourier Transform (FFT) \cite{fft-short}, Welch Power Spectral Density (PSD) \cite{Welch-psd-short}, the periodogram and multitaper \cite{multitaper-short}. These techniques collectively transform the ROI time series into frequency domain representations, providing a nuanced perspective on the underlying neural dynamics. \\

In this paper, we propose a new graph neural network, SpectBGNN, which takes spectral features from each Region of Interest (ROI) time series and brain connectivity as input and outputs the predicted MASC-2 total score. To further improve the performance, we added graph spectral polynomial filtering layers, which have high versatility and can help the network harmonize the spectral features across all nodes to extract useful spectra \cite{graph_filter-short}. To compare and evaluate the results, we tested our methodology on 70 autistic children and 26 matched unaffected subjects with 2 different GNNs: graph attention network (GAT) \cite{gat-short} and BrainGNN \cite{braingnn-short} with 5-fold cross-validation. 

\section{Method}
The general structure of our network, SpectBGNN, is shown in Fig.~\ref{fig}. The graph for the brain is defined by $G = (v,e)$ where $v = \{v_1, v_2, ... v_n\}$ (n = 268) represents each ROI in the brain. The edge $e_{ij}$ represents the connectivity between node $v_i$ and node $v_j$, which is quantified by the partial correlation between ROI time series ${x(t)_i}$ and ${x(t)_j}$. Here, we define vector $h_i = \{h_{i1}, h_{i2}, ... h_{ik}\}$ as the node features. \\

\subsection{Node spectral features}\label{node_spectral}
We applied multiple spectral analysis algorithms on the ROI time series to decode frequency domain information as node features. The first algorithm is the Fast Fourier Transform (FFT) \cite{fft-short}. For each node $v_i$, we first calculated the FFT for the ROI time series $x_i$ using the equation \eqref{eqn_1} to construct $h_i$ = $\{|X_1|, |X_2|, ..., |X_K|\}$ where $K$ is the number of the data points in the signal. The second method is to use the periodogram as the node features. So each node feature is constructed as $h_i = \{|X_1|^2, |X_2|^2, ..., |X_K|^2\}$.
\begin{equation} 
\label{eqn_1}
X_{k} = \sum^{K-1}_{m=0} x_i(m)e^{-i2\pi k m/n}  \quad  k = 0, .. K-1  
\end{equation} 
 
The third algorithm is Welch's Power Spectral Density (PSD), which estimates the power of the spectrum. Welch's PSD algorithm first divides the signal into successive blocks and applies the FFT algorithm to them. Finally, it takes the average of the squares of the FFT magnitudes \cite{Welch-psd-short}. In this case, the node feature $h_i = \{{\hat R}_1,{\hat R}_2, ... ,{\hat R}_N \}$, where ${\hat R}_n$ is calculated by equation \eqref{eqn_2}: 
\begin{equation} 
\label{eqn_2} 
{\hat R}_x(\omega_n) = \frac{1}{M}\sum_{m=0}^{M-1} FFT_k(x_m)^2 {X_m(\omega_n)^2}_m ,
\end{equation} 
where $M$ represents the number of the blocks and $x_m(n) = x_m(n + mN)$ with $ n = 0, 1, 2,...K-1$. 

The fourth algorithm is multitaper \cite{multitaper-short} which estimates the spectral density of the signal after tapering the signals with multiple preset sequences. The method conducts multiple element-wise multiplications of the signal with $K$ different Slepian sequences \cite{slepian-short} and each multiplication result is converted to the frequency domain by FFT. The final result is obtained by averaging all FFT results as shown in equation \eqref{eqn_3}, where $s_k(n)$ represents the k-th Slepian sequence.

\begin{equation} 
\label{eqn_3} 
S_{k}(f) =\frac{1}{K} \sum^{K-1}_{k=0} \Delta{t} | \sum_{n=0}^{N-1} s_k(n) x(n) e^{-j2\pi fn\Delta t}|
\end{equation} 

\subsection{Graph Spectral Filtering}
The node features are first processed by a multi-layer perceptron (MLP) and the node embeddings are then fed into the graph spectral polynomial filtering layer \cite{spectral_filtering-short} as the filtering layer could attenuate the node features to concentrate on the more important frequency band for the brain ROI. The polynomial filtering is calculated by equation \eqref{polynomial_filter_eqn}. The $g_l$ represents the polynomial filter basis and $\alpha_l$ denotes the coefficients. The Laplacian matrix of the graph, $\hat{L}$, is calculated by $\hat{L} = I - \hat{D}^{-\frac{1}{2}}\hat{A}\hat{D}^{-\frac{1}{2}}$. $\hat{A}$ denotes the adjacency matrix with self-added loops to the graph $\hat{A} = A + I$. $\hat{D}$ is the degree matrix of adjacency matrix $\hat{A}$. We applied this graph spectral filtering to mitigate the impact of the spectral feature artifacts. 

\begin{equation}
\label{polynomial_filter_eqn}
Z = \sum_{l=0}^{l}\alpha_lg_l(\hat{L})hW
\end{equation}

\section{Experiments and Results}
\subsection{Experimental Setup}
\noindent\textbf{Datasets}~~~~
We conducted experiments on 70 IQ and age-matched ASD children, ages 8 to 15 years old, and 26 unaffected controls, who completed resting-state fMRI scans, clinical characterization, and passed motion quality checks. Full-scale IQ was evaluated with the Differential Ability Scales-II (DAS-II) \cite{DAS2-short}. The mean IQ for the ASD population is 98.8 and the standard deviation is 20.54. Parent-rated measures of anxiety included the Multidimensional Anxiety Scale for Children, 2nd edition (MASC-2) \cite{MASC2-short}. The MASC-2 total score in this dataset ranges from 40 to 90 with a mean of 61.6. 

Imaging was performed using a Siemens MAGNETOM Tim Trio 3 Tesla scanner. The fMRI was co-registered with T1 images into the MNI standard space \cite{mni_standard-short} and denoised with ICA-AROMA \cite{AROMA-short}. fMRI was subsequently parcellated into 268 regions of interest (ROI) using the Shen-268 atlas \cite{Shen_268-short}. The representative ROI time series were computed by averaging signals from all voxels within each ROI.

\begin{table}[!t]
  \caption{The mean and standard deviation for MAE values from the 5-fold cross-validation experiments are summarized. The best performance feature for each model is bolded. \\}
  \centering
  \begin{tabular}{l c ccc c ccc}
  \toprule
 & {\textbf{Node Features}} & MAE$\ \downarrow$ \\
   \midrule
\multirow{1}{*}{\textbf{CPM}} & - & 15.10 $\pm$ 1.76 \\
\midrule
\multirow{5}{*}{\textbf{GAT}} & \textbf{Correlation} & 14.96 $\pm$ 1.12\\
&\textbf{FFT} & \textbf{14.12 $\pm$ 1.33} \\
&\textbf{PSD} &  {14.24 $\pm$ 1.63} \\
&\textbf{periodogram} &  {15.23 $\pm$ 1.45} \\
&\textbf{multitaper} &  {14.62 $\pm$ 1.17} \\
\midrule
\multirow{5}{*}{\textbf{BrainGNN}} &\textbf{Correlation}& 14.41 $\pm$ 1.54 \\
&\textbf{FFT}& \textbf{13.84 $\pm$ 1.35} \\
&\textbf{PSD}& 14.10 $\pm$ 1.43 \\
&\textbf{periodogram}& 14.10 $\pm$ 1.43 \\
&\textbf{multitaper}& 14.66 $\pm$ 1.09 \\
\midrule
\multirow{4}{*}{\textbf{SpectBGNN}} 
&\textbf{FFT}& \textbf{13.77 $\pm$ 1.35}  \\
&\textbf{PSD}& 13.83 $\pm$ 1.36 \\
&\textbf{periodogram}& 13.80 $\pm$ 1.33 \\
&\textbf{multitaper}& 13.81 $\pm$ 1.13 \\
  \bottomrule
\end{tabular}
\label{tab:experiment result}
\end{table}

\noindent\textbf{Implementation Details}~~~~
To train the network, we set the learning rate as 0.005 and run a total of 100 epochs. The learning rate is reduced to half for every 8 epochs. We applied mean squared error as the loss function and Adam Optimizer to train the network.

\begin{table*}[!t]
\caption{Quantitative comparison between (1) spectral features and correlations and (2) SpectBGNN and BrainGNN. Two-way repeated ANOVA tests are conducted to separate the variance caused by graph model and node feature choice. Row 1 through 5 are the ANOVA setup of the 2 factors and comparison of performance among Spectral features and Correlation on GAT, BrainGNN, and SpectBGNN. Row 6 is the ANOVA setup and results comparing the significance of the presence of the graph spectral filter. P-values for factor-2 results are calculated and p-values $<$ 0.05 are bolded.}
\centering
\begin{tabular}{c c c}
\toprule
\textbf{Factor 1} 
&\textbf{Factor 2} 
& \textbf{Factor 2 p-values} \\ 
\midrule
\text{graph models} & FFT, Correlation & \textbf{0.004}\\
\text{graph models} & PSD, Correlation & \textbf{0.016}\\
\text{graph models} & FFT, PSD & 0.189 \\
\text{graph models} & FFT, multitaper & 0.074 \\
\text{graph models} & FFT, periodogram  & 0.10 \\
{\text{FFT, PSD, periodogram, multitaper}} & {with or w/o spectral filtering layer} 
&\textbf{0.005}\\

\bottomrule
\end{tabular}
\label{tab:fft_vs_correlation and spectral_filters}
\end{table*}


\noindent\textbf{Baselines and Ablation Studies}~~~~
 Our baseline non-deep learning comparison method is Connectome-based Predictive Modeling (CPM) with ridge regression \cite{cpm-short}. The CPM hyperparameter alpha values, which is the coefficient for the regularization term, are searched from 0.5 to 5x$10^9$ and the best alpha value is picked by evaluating the model on the validation dataset. Baseline GNN models compared include GAT and BrainGNN. For each baseline GNN, we tested using Pearson's correlation or each of the spectral decomposition results as the input node features described in Sec.~\ref{node_spectral}. We also tested our SpectBGNN model using each of the spectral decomposition features as input.

 \noindent\textbf{Evaluation Methods}~~~~
 To achieve more reproducible results, we evaluated the networks' performances with 5-fold cross-validation and compared the Mean Absolute Error (MAE). We first split the data into 5 folds. Each time, we used one fold as the test set, another fold as the validation set, and all the other three as the training set. The network with the best performance in the validation set will be tested. All models are trained, validated, and tested with the same data split. To study the impact of spectral features, we carried out a 2-way ANOVA with repeated measures to calculate the statistical significance with one factor being the different graph-based models, and the other factor being node feature inputs. We chose repeated measures because for each fold we have repeated measurements as the data split is the same for all models. We also tested the effectiveness of the spectral features by comparing the performance of BrainGNN and SpectBGNN with spectral features as inputs. The 2-way ANOVA with repeated measures was set up with one factor being the presence of the graph spectral filtering layer (SpectBGNN or BrainGNN) and the other factor being the choice of spectral decomposition method. 

\subsection{Experimental Results} 
All mean and standard deviation of MASC-2 total score prediction errors are summarized in Table \ref{tab:experiment result}. Most graph-based models show lower MAEs than the non-deep learning CPM baseline averaging 5-fold cross-validation. For all graph-based methods, we find that using FFT or PSD as node features shows higher average performance compared to using Pearson's correlation as node features with p = 0.004 and p = 0.016, respectively. (Table \ref{tab:fft_vs_correlation and spectral_filters}, Row 1 and 2). Even though FFT shows a lower average MAE across all models, ANOVA results did not show FFT is significantly different from all other spectral features (Table \ref{tab:fft_vs_correlation and spectral_filters}, Row 3 to 5). 

To test the efficiency of the graph spectral filter layer on the spectral features, we compared the performance between BrainGNN and SpectBGNN with different spectral features as input. The results  showed significant improvement in performance  (p = 0.03) when the graph spectral filter was integrated (Table \ref{tab:fft_vs_correlation and spectral_filters}, Row 6).

\section{Discussion and Conclusions}
In this paper, we proposed SpectBGNN, a new graph neural network, and applied it to fMRI data of 26 neurotypical and 70 ASD children to predict their MASC-2 total score. The network takes spectral features as node features and integrates a graph spectral filtering layer. Our performance is in line with another relevant study on the regression of other cognitive scores with state-of-the-art graph models~\cite{IQ_prediction-short}.  
We found that using spectral features as input achieves comparable or even better results than using Pearson's correlation in graph-based methods. The results did not show statistically significant differences between using FFT and other spectral features. This could be explained by the similarity of the feature computation methods. We also showed that polynomial graph spectral filtering increases the network's performance. However, one of the limitations of this study pertains to the relatively small size of the dataset utilized. The constraints imposed by the limited number of observations may impact the robustness and generalizability of our findings. 
These results suggest the potential advantage of utilizing spectral features as node features for graph-based methods. Furthermore, adjustments to the network, such as integrating the graph spectral filtering layers, need to be made to release the potential of spectral features.

\section{Compliance with Ethical Standards}
The data acquisition was reviewed and approved by the local ethical committee~(institutional review board at the Yale University School of Medicine). Each participant’s parent provided informed consent according to the institutional review board at the Yale University School of Medicine. Each child provided verbal and written consent. 
\section{Acknowledgments}
\label{sec:acknowledgments}

The authors would like to thank all participants. This study is supported by the National Institute of Neurological Disorders and Stroke~(NINDS) of the National Institutes of Health through grant R01~NS035193 and National Institute of Child Health and Human  Development~(NICHD)  grant  R01~HD083881.

\bibliographystyle{IEEEbib}
\bibliography{ISBI2024_ASD}

\end{document}